# Jahn-Teller Distortion and Multiple-spin-state Analysis of Single-atom Vacancy in Graphene-nano-ribbon

Norio Ota

Graduate School of Pure and Applied Sciences, University of Tsukuba,

1-1-1 Tenoudai Tsukuba-city 305-8571, Japan

A single-atom vacancy defect and its array in graphene and graphite were considered to be one candidate carrying the room-temperature ferromagnetism. Applying density functional theory to a single-atom vacancy in graphene-nano-ribbon (GNR), a detailed relationship between the multiple-spin-state and the Jahn-Teller distortion was studied. An equilateral triangle of an initial vacancy having six unpaired electrons had distorted to isosceles triangle by the Jahn-Teller effect. Among capable spin-state of Sz=6/2, 4/2 and 2/2, the most stable one was Sz=2/2. Total energy was 15.6 kcal/mol lower (stable) than that of the initial one and a sum of spin density (magnetic moment) around one vacancy was 1.49 µB. Amazing result was obtained in case of Sz=4/2. Initial flat ribbon turned to three dimensionally curled one. There appears ferromagnetic spin distribution on GNR. Total energy was -15.5kcal/mol, which was very close to that of Sz=2/2. Such calculation suggested the coexistence of flat ribbon and curled ribbon by generating vacancies. Bi-layered AB stacked GNR was analyzed in case of α-site vacancy and also β-site one. The most stable spin state was Sz=2/2 in both cases. These distorted vacancy triangle show 60 degree clockwise rotation from beta- to alpha-site, which is consistent with several experimental observations by using a scanning tunneling microscope.

## 1. Introduction

These ten years, carbon based room-temperature ferromagnetic materials are experimentally reported[1)-6)]. They are graphite and graphene like materials. Such peculiar ferromagnetic new materials would greatly enhance the application of spintronics technology. Such experimental evidence of magnetic ordering were explained by the presence of impurities[7)], edge irregularities[8)-10)] or defects[11)-16)]. Single-atom vacancy defect is important to understand the basic mechanism of magnetic behavior. In experiments, in order to create vacancy defects, high energy particles were irradiated to materials[11)-13)]. Observation of single-atom vacancy was eagerly done by a scanning tunneling microscope method[14)-16)]. Unfortunately, there are little information on magnetic behaviors and spin distributions in atomic scale. Theoretical calculations predicted the importance of Jahn-Teller distortion [17)-18)]. However, there are little explanation on a detailed relationship between the multiple-spin-state and the Jahn-Teller distortion. There are six unpaired electrons around one single-atom vacancy, which suggests a capability of multiple-spin-state as like Sz=6/2, 4/2, and 2/2. We need detailed spin-state calculations and to find the most stable spin-state accompanying with Jahn-Teller distortion.

This paper focuses on a relationship between the Jahn-Teller distortion and the multiple-spin-state of single-atom vacancy in graphene-nano-ribbon (GNR). In a final section, we will report the graphite surface like model with an AB stacked bi-layer GNR.

## 2. Calculation Method

We have to obtain the (i) spin density, (ii) total energy, and (iii) optimized atom configuration depending on a respective given spin state Sz to clarify magnetism. Density functional theory (DFT) [19)-20)] based generalized gradient approximation (GGA-PBEPBE) [21)] was applied utilizing Gaussian03 package[22)] with an atomic orbital 6-31Gd basis set[23)]. In this paper, total charge of model super-cell is set to be completely zero. Inside of a super-cell, three dimensional DFT calculation was done. One dimensional periodic boundary condition was applied to realize an unlimited length graphene-ribbon.

Self-consistent energy, atomic configuration and spin density calculations are repeated until to meet convergence criteria. The required convergence on the root mean square density matrix was less than 10E-8 within 128 cycles.

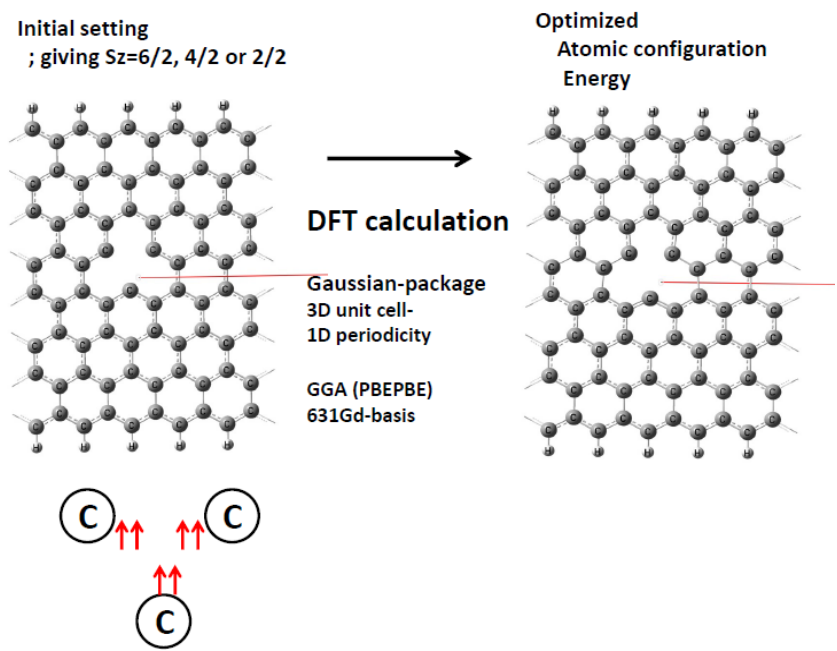

**Fig.1** Initial setting of single-atom vacancy in mono-layer graphene-nano-ribbon in left, where six unpaired electrons carry the multiple-spin-state and the Jahn-Teller distortion. Gaussian package DFT calculations gave the optimized atomic configuration and spin density.

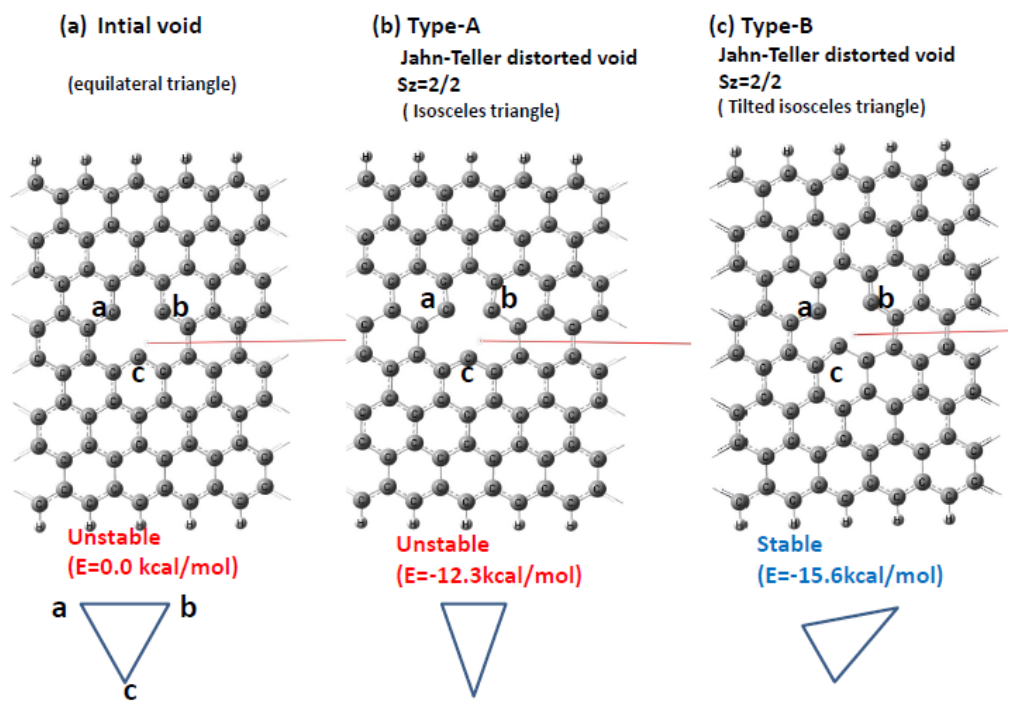

**Fig.2** Jahn-Teller distorted vacancy was grouped as Type-A and Type-B. Type-A has an isosceles triangle vacancy perpendicular to GNR, whereas Type-B tilted to GNR.

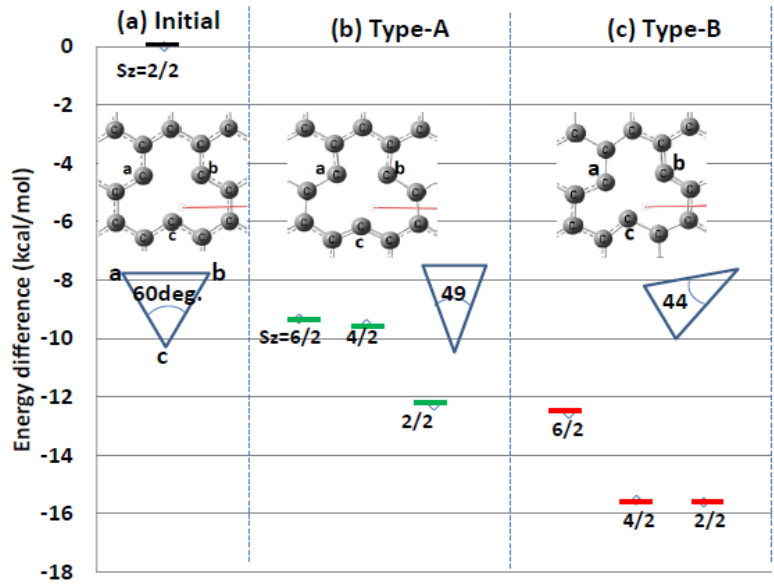

**Fig.3** Total energy decreases with decreasing spin-state Sz. Initial vacancy triangle (a) was equilateral with minimum angle of 60°. In Type-A (b), distorted isosceles triangle has an angle of 49°. More distorted Type-B show the lowest (stable) total energy with a smaller angle of 44°.

**Table 1,** DFT calculation results of graphene-nano-ribbon with a single-atom vacancy. The lowest energy (stable) case was B3 with Sz=2/2. In case of B2 (Sz=4/2), energy value was very close to B3. Configuration turned from a flat ribbon to a three dimensionally curled ribbon.

| Distorted type | none | Type A | | | Type B | | |
|---|---|---|---|---|---|---|---|
| Case number | A0 | A1 | A2 | A3 | B1 | B2 | B3 |
| Given Sz | 2/2 | 6/2 | 4/2 | 2/2 | 6/2 | 4/2 | 2/2 |
| Energy difference (kcal/mol/unit cell) | 0 | -9.33 | -9.51 | -12.31 | -12.58 | -15.54 | -15.62 |
| Ribbon configuration | Flat | Flat | Flat | Flat | Flat | Curled | Flat |
| Triangle distance ab | 2.48 | 2.15 | 2.11 | 2.14 | 2.59 | 2.61 | 2.59 |
| (A)        bc | 2.48 | 2.6 | 2.59 | 2.6 | 2.61 | 2.64 | 2.61 |
| ca | 2.48 | 2.6 | 2.59 | 2.6 | 1.96 | 1.82 | 1.92 |
| Smallest angle (deg.) | 60 | 48.8 | 48.1 | 48.6 | 44.3 | 40.6 | 43.5 |
| Mulliken charge (e) a | 0.21 | 0.25 | 0.25 | 0.25 | 0.25 | 0.22 | 0.25 |
| b | 0.21 | 0.25 | 0.25 | 0.25 | 0 | 0 | -0.01 |
| c | 0.22 | 0.05 | 0.03 | 0.04 | 0.27 | 0.23 | 0.27 |
| Spin density ($\mu_B$) a | -0.08 | 0.24 | 0.03 | 0.21 | 0.19 | 0.01 | 0.15 |
| b | 0.77 | 0.24 | 0.03 | 0.21 | 1.09 | 0.79 | 1.06 |
| c | 0.97 | 1.15 | 0.81 | 1.12 | 0.34 | -0.17 | 0.28 |
| a+b+c | 1.66 | 1.63 | 0.87 | 1.54 | 1.62 | 0.63 | 1.49 |
| Ribbon magnetism | Ferrimag. | Ferrimag. | Ferromag. | Ferrimag. | Ferrimag. | Ferromag. | Ferrimag. |

### 3, Initial model of Graphene-nano-ribbon with a single vacancy

An initial calculation model of graphene–nano-ribbon with a single carbon atom vacancy is shown in the left hand side of Fig.1. Super-cell was [$C_{79}H_{10}$], where upper and lower zigzag edge carbons were all hydrogenated. Ribbon width was 17.80 Å and one dimensional super-cell length was 12.41 Å. A center positioned carbon was removed to make one vacancy defect. Around this vacancy, there are three carbons, which make an equilateral triangle with an edge length of 2.48 Å. These three radical carbons bring six unpaired electrons. In order to clarify magnetic characteristics, we should consider these six electrons interaction, which mean to study a detailed multiple-spin-state analysis.

### 4, Multiple-spin-state

Six unpaired electrons around one vacancy are capable to set multiple-spin-state parameter to be $Sz$=6/2, 4/2, or 2/2. For DFT calculation, initial atom positions and spin parameter $Sz$ were installed. After repeating self-consistent calculation, total energy, optimized atomic configuration, charge and spin density were obtained as like Fig. 2 (b) and (c), where initial equilateral triangle vacancy turned to an isosceles one. In Fig.2, there were two types of isosceles triangle. Type-A was perpendicular to a ribbon axis (a red line), whereas Type-B tilted. Such distortion originates from Jahn-Teller effect as discussed by Yazyev and Helm[17]. Here, more detailed relationship between the multiple-spin-state and energy stability was calculated. by grouping into Type-A and –B. Results were shown in Fig.3 and Table 1. Initial equilateral triangle vacancy energy level was defined to be zero. In Type-A, the most stable spin state was $Sz$=2/2. Energy level reduced to -12.31 kcal/mol, where triangle's smallest angle was 49°. All of Type-B spin state show more remarkable energy decrease. In case of $Sz$=2/2, energy difference became -15.62 kcal/mol. The reason is more serious triangle distortion with a sharper angle of 44°. Distance between carbon atom a and c was shortened to 1.92 Å suggesting a sigma like bonding between those two. Magnetic moment (a sum of spin density) around a vacancy was 1.49 $\mu_B$, which coincides very well with the case of smaller defect concentration calculation using the SIESTA code by Yazyev and Helm[17].

Spin-densities of Type-A and –B were illustrated in Fig.4. In both cases, we can see a large up-spin cloud at a vacancy triangle apex carbon site, which is composed by both sigma and pi-component of electron orbits belonged to a carbon atom. Such spin cloud was familiar to an edge positioned radical carbon[10]. Inside of zigzag edge GNR, except vacancy region, there appears up- and down-spin clouds alternately arranged one by one regularly observed in magnetically polarized GNR[9),10)].

### 5, Curled ribbon

Amazing result was obtained in case of $Sz$=4/2 as illustrated in Fig.5, where the shape of ribbon itself was curled three dimensionally as seen in a tilt view and side view. A single vacancy triangle was also three dimensionally distorted. The smallest angle of isosceles triangle was 40.6° in case of $Sz$=4/2, whereas 43.5° in $Sz$=2/2. Additional interesting feature was figured in Fig.6. Spin distribution of (d) Type-B $Sz$=4/2 show a ferromagnetic feature, that is, both of upper and lower zigzag edge carbons carry up-spin clouds (red colored). Total energy of (d) was -15.5 kcal/mol which was very close value to that of Type-B $Sz$=2/2 (-15.6 kcal/mol). Such calculation suggested the coexistence of flat ribbon and curled ribbon with vacancies created by high energy electron and atom irradiation.

### 6. Bi-layer graphene-nano-ribbon

Bi-layered AB stacked GNR was analyzed in case of α-site vacancy and also β-site one. In Fig.7, yellow marked atoms are surface (first) layer including a single-atom vacancy, whereas gray atoms are back (second) layer without any vacancy. Unit cell is [$C_{47}H_6$-$C_{48}H_6$]. In case of α-site vacancy with a triangle in a first layer, we can look into one center carbon belonged to a second layer. On the contrary, in case of β-site one, we cannot look into any carbon inside of a triangle vacancy. Distance between two layers was 4.09~4.13 Å as summarized in Table 2. In bi-layer GNR too, distorted triangles were classified to Type-A and –B as shown in Fig.8. In every cases, the most stable spin state was $Sz$=2/2. Total energy of Type-B was lower ( more stable) than that of Type-A. Distorted isosceles triangle of Type-B show that the smallest angle was 42° smaller than that of 53° in Type-A. It should be noted that in Type-B both α-site vacancy and β-site one resulted almost the similar energy (see Table 2). Notable result is the clockwise angle of a vacancy triangle defined in the figure in lower part of Table 2. The clockwise angle of α-site vacancy triangle θ was 300°, whereas β-site 240°. There appear 60 degree rotation from β-site to α-site vacancy triangle. Surprisingly, such calculation result was consistent with several experimental observations by using a scanning tunneling microscope[14),16)]. In Fig.9, spin density configuration of AB-stacked bi-layer GNR was illustrated. Spin density appeared only in the surface (first) layer. Back (second) layer show no spin density. We can see fruit pear like up-spin cloud at an apex carbon site of a triangle. Around a vacancy, summed spin-density (magnetic moment) was 1.37$\mu_B$.

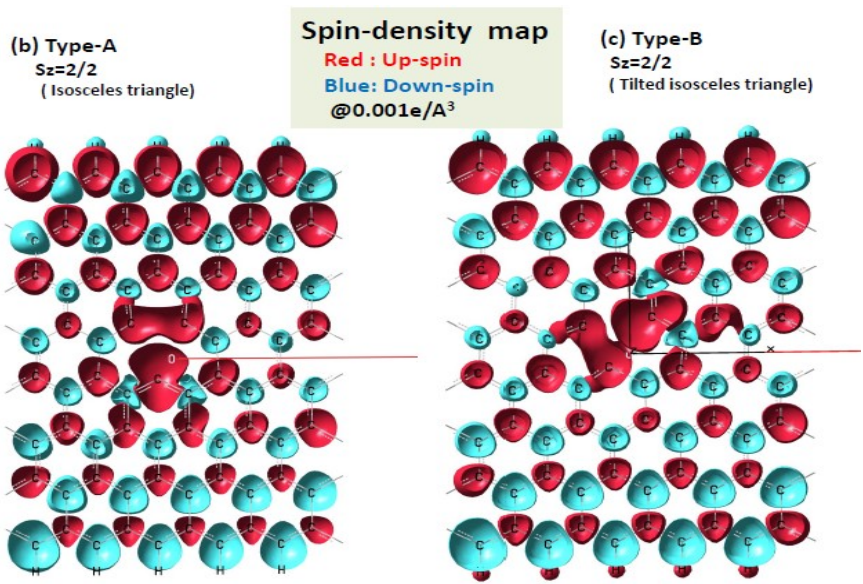

**Fig.4** Spin-density distribution of Type-A and Type-B. A fruits pear resembled up-spin cloud was observed at an apex site of vacancy triangle. Inside of GNR, there exist alternately arranged up- and down-spin clouds one by one.

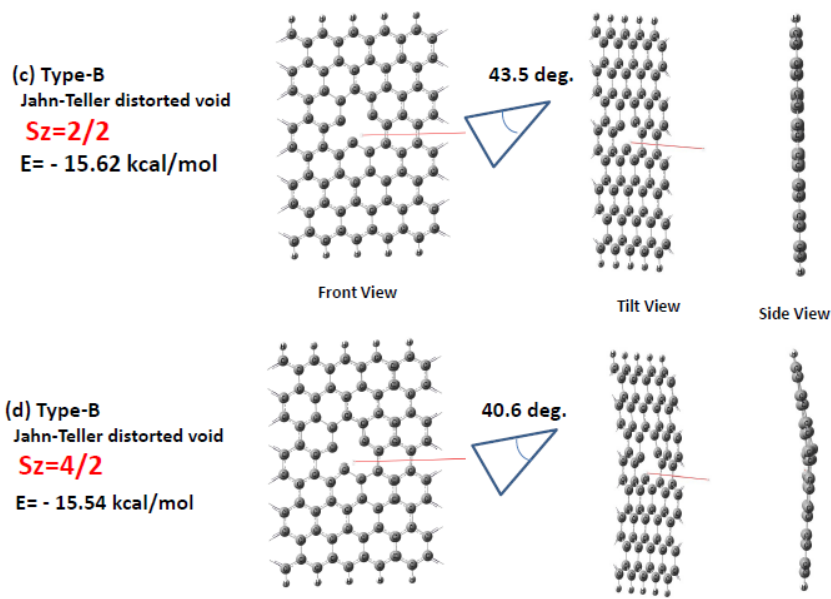

**Fig.5** In case of Sz=4/2, ribbon shape was three dimensionally curled in Type-B. Distorted vacancy triangle had a sharper angle of 40.6 deg. This suggests a capability of coexistence of flat and curved GNR.

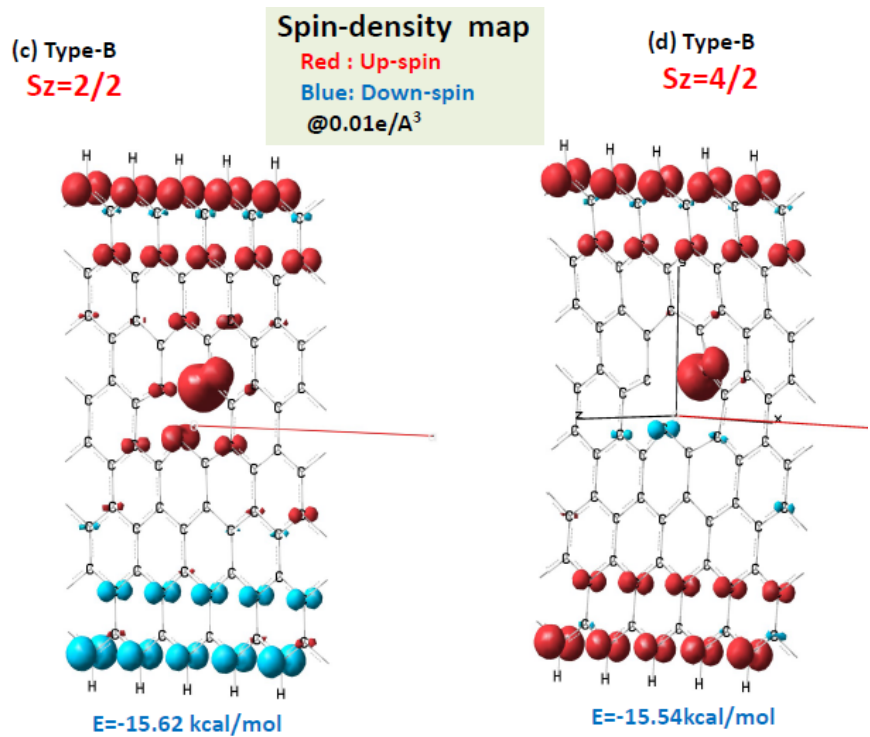

**Fig.6** Comparison of spin density configuration of Type-B Sz=2/2 (c) and Sz=4/2 (d). Curved ribbon structured high-spin-state (d) show remarkable ferromagnetic-like spin polarization at both ends of zigzag-edge carbons.

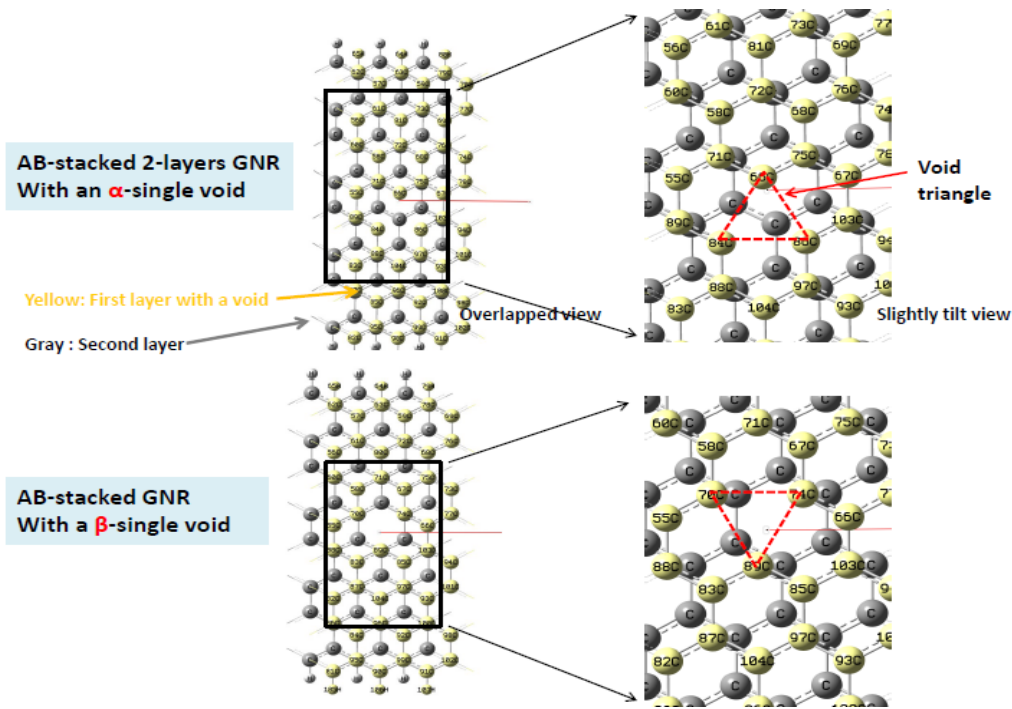

**Fig.7** AB-stacked bi-layer GNR with a single-atom vacancy at an alpha-site (upper figure) and a beta-site (lower one). Yellow marked balls are first layer atoms, whereas gray balls second (back) layer one.

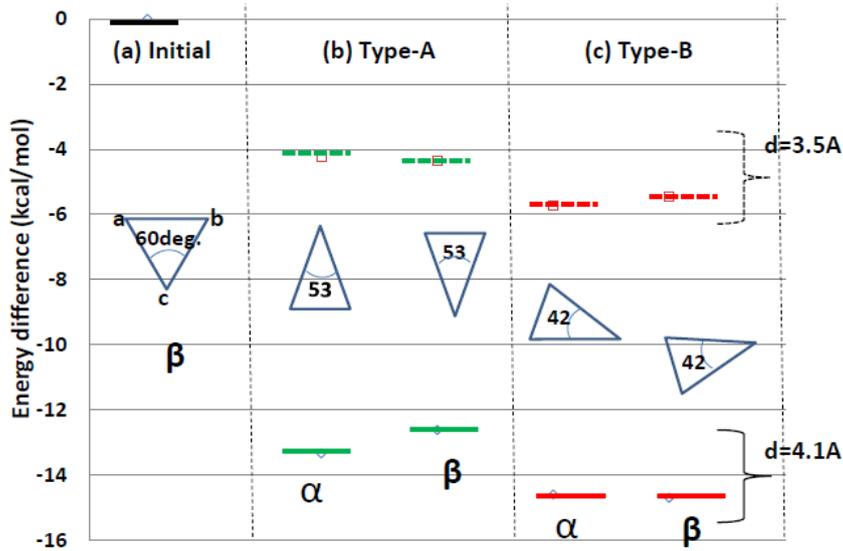

**Fig.8** Energy and vacancy triangle were compared in AB-stacked bi-layer GNR. Type-B distorted alpha- and beta-site vacancy resulted similar total energy at an optimized stacked distance around 4.1Å.

**Table 2** DFT calculation results of bi-layer GNR with a single-atom vacancy. Most stable Jahn-Teller distorted case was alpha-B and beta-B. The clockwise angle $\theta$ of αB was 300°, while βB 240°. There appear 60 degree rotation from β-site to α-site vacancy triangle. Such calculation result was consistent with several experimental observations by using a scanning tunneling microscope[14)16)].

| Case | Initial | αA | αB | βA | βB |
|---|---|---|---|---|---|
| Void site | β | α | α | β | β |
| Distorted type | none | Type A | Type B | Type A | Type B |
| Given Sz | 2/2 | 2/2 | 2/2 | 2/2 | 2/2 |
| optimized stacked distance d (A) | 4.1 | 4.09 | 4.11 | 4.13 | 4.11 |
| Energy difference (kcal/mol/unit cell) | 0 | -13.33 | -14.59 | -12.61 | -14.69 |
| Ribbon configuration | Flat | Flat | Flat | Flat | Flat |
| Triangle distance ab (A) | 2.45 | 2.63 | 2.64 | 2.34 | 2.53 |
| bc | 2.49 | 2.34 | 2.53 | 2.63 | 2.64 |
| ca | 2.49 | 2.63 | 1.84 | 2.63 | 1.84 |
| Smallest angle (deg.) | 60 | 52.8 | 41.7 | 52.8 | 41.7 |
| Direction of triangle θ(deg.) | 0 | 180 | 300 | 0 | 240 |
| Mulliken charge (e) a | 0.22 | 0.04 | 0.21 | 0.21 | 0.24 |
| b | 0.22 | 0.21 | -0.01 | 0.21 | -0.01 |
| c | 0.17 | 0.21 | 0.24 | 0.04 | 0.21 |
| Spin density ($\mu_B$) a | 0.33 | 0.9 | 0.04 | 0.25 | 0.25 |
| b | 0.34 | 0.25 | 1.08 | 0.25 | 1.08 |
| c | 0.84 | 0.25 | 0.25 | 0.9 | 0.04 |
| a+b+c | 1.51 | 1.4 | 1.37 | 1.4 | 1.37 |
| Ribbon magnetism | Ferrimag. | Ferrimag. | Ferrimag. | Ferrimag. | Ferrimag. |
| Void triangle | | | | | |

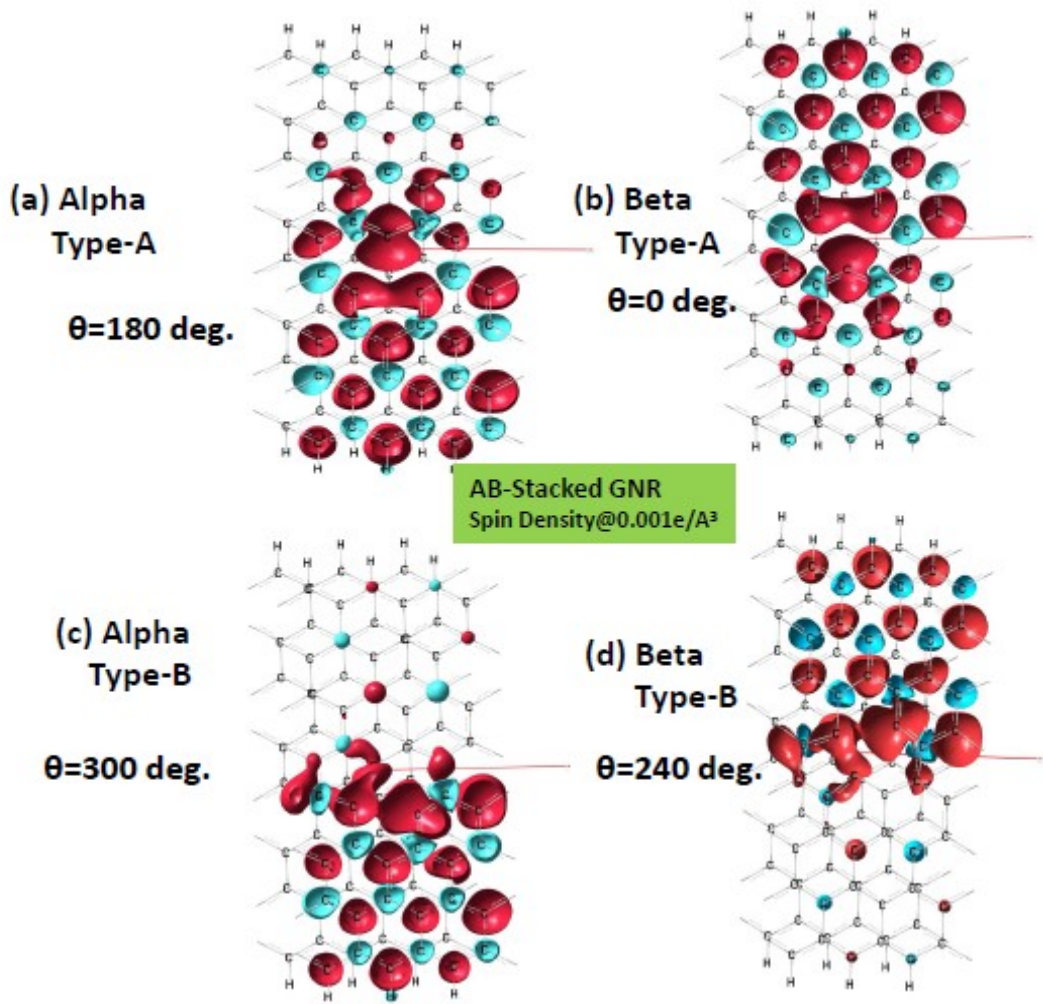

**Fig.9** Spin density configuration of AB-stacked bi-layer GNR is compared, where red cloud show up-spin and blue one down-spin. Spin density appeared only on the surface (first) layer. Back (second) layer show no spin density. We can see a fruit pear like up-spin cloud at an apex carbon site of a vacancy triangle.

## 7. Conclusion

Applying density functional theory to a single-atom vacancy in graphene-nano-ribbon (GNR), a relationship between the multiple-spin-state and the Jahn-Teller distortion was studied. An equilateral triangle of an initial vacancy had distorted to isosceles triangle by the Jahn-Teller effect for every spin states $Sz=6/2$, $4/2$ and $2/2$. The most stable spin state was $Sz=2/2$. Sum of spin density (magnetic moment) around one vacancy was 1.49 $\mu_B$. Amazing result was obtained in case of $Sz=4/2$, where the shape of ribbon itself was three dimensionally curled and there appear ferromagnetic spin distribution on GNR. Bi-layered AB stacked GNR was analyzed in case of α-site vacancy and also β-site one. The most stable spin state was $Sz=2/2$ in both cases. These distorted vacancy triangles presented 60 degree clockwise rotation from β-site to α-site. This calculation result is consistent with several experimental observations.

Preprint   first version:   Aug. 26, 2014
By Norio Ota